\documentclass[twocolumn,preprintnumbers,floats,amsmath,amssymb,aps,prb,showpacs]{revtex4}

\usepackage{graphics}
\usepackage{amsmath}
\usepackage{dcolumn}
\usepackage{epsfig}
\begin{document}

\preprint{(submitted to Phys. Rev. B)}

\title{Alloy surface segregation in reactive environments:\\
A first-principles atomistic thermodynamics study\\
of Ag$_3$Pd(111) in oxygen atmospheres}

\author{John R. Kitchin}
\altaffiliation[Permanent address:]{Carnegie Mellon University,
  Department of Chemical Engineering, 5000 Forbes Ave., Pittsburgh, PA
  15213 (USA)}

\author{Karsten Reuter}
\author{Matthias Scheffler}

\affiliation{Fritz-Haber-Institut der Max-Planck-Gesellschaft
Faradayweg 4-6, 14195 Berlin (Germany)}

\date{\today}

\begin{abstract}
We present a first-principles atomistic thermodynamics framework to describe the structure, composition and segregation profile of an alloy surface in contact with a (reactive) environment. The method is illustrated with the application to a Ag$_3$Pd(111) surface in an oxygen atmosphere, and we analyze trends in segregation, adsorption and surface free energies. We observe a wide range of oxygen adsorption energies on the various alloy surface configurations, including binding that is stronger than on a Pd(111) surface and weaker than that on a Ag(111) surface. This and the consideration of even small amounts of non-stoichiometries in the ordered bulk alloy are found to be crucial to accurately model the Pd surface segregation occurring in increasingly O-rich gas phases.
\end{abstract}



\maketitle

\section{Introduction}

The chemical properties of an alloy surface like its catalytic function or corrosion resistance depend intricately on the detailed surface composition. This compositional, and concomitant geometric and electronic structure is not only different from the corresponding surfaces of the parent metals, but differs often also substantially from the bulk alloy due to segregation of one species to the surface. The resulting segregation profile can depend on the surface orientation, the bulk composition, and temperature. A reactive environment, such as one where catalysis or corrosion occurs, can further modify the chemical composition of the surface, if one alloy component interacts more strongly with a gas phase species than the others, resulting in adsorbate-induced segregation. It is consequently usually not possible to predict the properties and functions of alloy surfaces in materials science applications solely on the basis of the formal bulk composition; the effects of the corresponding (possibly reactive) environment must explicitly be considered.

Previous theoretical attempts to address the segregation thermodynamics at an alloy surface from first-principles have mostly focused on the interdependence with the bulk alloy reservoir. This includes tabulated segregation energies of a transition metal impurity in another transition metal \cite{ruban99a,ruban99b}, or approaches based on the coherent potential approximation \cite{ropo05}, screened generalized perturbation method \cite{pourovskii01} or surface cluster expansions \cite{drautz01,muller03,wieckhorst04,fahnle05}. The understanding emerging from these studies points at a crucial influence of the bulk reservoir structure and composition on the surface segregation profile, even if this concerns only slight excesses of one type of atom in samples with equal nominal bulk composition. This complexity has not yet been met in the scarce attempts to also explicitly account for the effect of a (reactive) environment \cite{christofferssen02,han05,piccinin08}, but the results reveal nevertheless an at least equally sensitive dependence of the adsorbate-induced segregation on the detailed adsorption properties and gas phase conditions. 

In view of these findings we set up a first-principles atomistic thermodynamics \cite{weinert86,scheffler88,kaxiras87,qian88,reuter02a} framework based on density-functional theory that is able to address the structure and composition of an alloy surface while quantitatively accounting for the dependencies on both the bulk alloy and the gas phase reservoir. The method is applied to obtain the segregation profile of a Ag$_3$Pd(111) surface in contact with an oxygen atmosphere, and we illustrate the intricate role of the bulk alloy reservoir by discussing the effect of already a small amount of non-stoichiometries (in form of anti-sites) in a nominally ordered Ag$_3$Pd bulk alloy. Not surprisingly we find Ag-segregation to occur under O-poor environmental conditions, e.g. typical for ultra-high vacuum experiments, while Pd-segregation is preferred with increasing oxygen content in the gas phase. Disentangling the various factors contributing to this segregation profile reveals, however, an intriguing complexity that needs to be captured in the modeling to obtain quantitatively accurate results. In the course of this analysis we can thus scrutinize various approximate treatments suggested in the literature, like the neglect of non-stoichiometries in the second or deeper substrate layers, the substitution of bulk segregation with that of dilute impurities in the parent metal hosts, or the interpolation of bonding properties at the alloy surface from those known at the parent metal surfaces.

\section{Theory}

\subsection{Surface free energy}

In the following we describe a first-principles atomistic thermodynamics framework to address alloy surface segregation in reactive environments. Our modeling aims at a system composed of a gas phase containing species $j$ at partial pressures $p_j$ and a common temperature $T$, and a multi-component alloy $A_{n_{A({\rm bulk})}}B_{n_{B({\rm bulk})}}\dots I_{n_{i({\rm bulk})}}$, where $x_{i({\rm bulk})}= n_{i({\rm bulk})}/\sum_i n_{i({\rm bulk})}$ is the mole fraction of species $i$ $(i=A,B,...,I)$ in the alloy bulk. To evaluate the surface free energy of a particular surface of this alloy in the given environment, we can concentrate on the part of the system that is affected by the surface, which contains (normalized to surface area $A_{\rm surf}$) $N_j$ gas phase particles and a total of $N$ solid particles at $x_i$ mole fractions. The most stable surface structure and composition is then the one, which minimizes the surface free energy, defined as
\begin{eqnarray}
\label{general_surf}
\nonumber
\gamma(T,\{p_j\},\{N_j\},N,\{x_i\}) \;= \\ \nonumber
 \frac{1}{A_{\rm surf}} \; \Bigg[ \;
G(T,\{p_j\},\{N_j\},N,\{x_i\}) \; - \\ \nonumber
 \left. \sum_{i} \left( N x_i \right) \mu_i - \sum_{j} N_j \mu_j \right] 
\quad , 
\end{eqnarray}
where $G(T,p_j,N_j,N,x_i)$ is the Gibbs free energy of the surface, and $\mu_i$ and $\mu_j$ are the chemical potentials of alloy species $i$ and gas phase species $j$, respectively.

For clarity, we will develop the methodological framework in this work for a binary alloy in contact with a single component gas phase, with obvious generalizations to more component systems. Specifically, we consider a ${\rm Ag}_{n_{Ag({\rm bulk})}}{\rm Pd}_{n_{Pd({\rm bulk})}}$ alloy and an oxygen atmosphere described by an oxygen pressure $p_{\rm O_2}$. In view of the total energy calculations described below, we will furthermore consider the surface affected part of the alloy to be described by an inversion-symmetric slab geometry exposing two identical surfaces. With $A_{\rm surf}$ then denoting the area of the surface unit-cell, eq. (\ref{general_surf}) simplifies to
\begin{eqnarray}
\lefteqn{\gamma(T,p_{\rm O_2},N_{\rm O},N_{\rm slab},x_{\rm Ag},x_{\rm Pd}) \;=} \\ \nonumber
\frac{1}{2 A_{\rm surf}} \; \bigg[ && \!\!\!\!\!\!\!
G_{\rm slab}(T,p_{\rm O_2},N_{\rm O},N_{\rm slab},x_{\rm Ag},x_{\rm Pd}) \; - \\ \nonumber
&& \!\!\!\!\!\!\! N_{\rm slab} x_{\rm Ag} \mu_{\rm Ag} - N_{\rm slab} x_{\rm Pd} \mu_{\rm Pd} - N_{\rm O} \mu_{\rm O} \; \bigg] \; =\\ \nonumber
\; \frac{1}{2 A_{\rm surf}} \; \bigg[ && \!\!\!\!\!\!\!
G_{\rm slab}(T,p_{\rm O_2},N_{\rm O},N_{\rm slab},x_{\rm Ag},x_{\rm Pd}) \; - \\ \nonumber
&& \!\!\!\!\!\!\! N_{\rm slab} x_{\rm Ag} (\mu_{\rm Ag} - \mu_{\rm Pd}) - N_{\rm slab} \mu_{\rm Pd} - N_{\rm O} \mu_{\rm O} \; \bigg] \quad ,
\label{surf_specific}
\end{eqnarray}
where $N_{\rm slab}$ refers now to the total number of metal atoms per surface unit-cell in the slab, $N_{\rm O}$ similarly to the number of O atoms in the surface geometry, and in the second step we have exploited that the mole fractions sum to unity by construction ($x_{\rm Ag} + x_{\rm Pd} = 1$). 

Within our thermodynamic theory we assume the surface to be in equilibrium with the surrounding gas phase, and correspondingly the oxygen chemical potential is determined in the entire system by the gas phase reservoir, $\mu_{\rm O} = 1/2 \mu_{\rm O_2(gas)}(T,p_{\rm O_2})$. The alloy surface is furthermore assumed to be in equilibrium with the underlying bulk alloy reservoir. This means that the chemical potentials $\mu_{\rm Ag}$ and $\mu_{\rm Pd}$ are not independent either, but are related by the bulk alloy chemical potential $\mu_{\rm bulk}$, i.e. the Gibbs free energy per atom of the bulk alloy. For a bulk alloy of composition ${\rm Ag}_{n_{\rm Ag (bulk)}}{\rm Pd}_{n_{\rm Pd( bulk)}}$, the relationship between the chemical potentials is therefore
\begin{equation}
\mu_{\rm bulk} \;=\; x_{\rm Ag(bulk)} (\mu_{\rm Ag} - \mu_{\rm Pd}) +  \mu_{\rm Pd} \quad , 
\label{mu_bulk}
\end{equation}
where we have again used the constraint on mole fractions. With this equation and introducing the difference in chemical potentials of the two metal species, $\Delta \mu_{\rm Ag-Pd} = \mu_{\rm Ag} - \mu_{\rm Pd}$, we can rewrite eq. (\ref{surf_specific}) to
\begin{eqnarray}
\label{general_result}
\lefteqn{\gamma(T,p_{\rm O_2},N_{\rm O},N_{\rm slab},x_{\rm Ag},x_{\rm Pd}) \;=\; \frac{1}{2 A_{\rm surf}}} \\ \nonumber
\bigg[ && \!\!\!\!\!\!\!
G_{\rm slab}(T,p_{\rm O_2},N_{\rm O},N_{\rm slab},x_{\rm Ag},x_{\rm Pd}) \; - \\ \nonumber
&& \!\!\!\!\!\!\! N_{\rm slab} \mu_{\rm bulk} -
N_{\rm slab} \Delta \mu_{\rm Ag-Pd} \left(x_{\rm Ag} - x_{\rm Ag(bulk)}\right) \; - \\ \nonumber
&& \!\!\!\!\!\!\! – \frac{N_{\rm O}}{2} \mu_{\rm O_2(gas)}(T,p_{\rm O_2})
\bigg] \quad .
\end{eqnarray}
This equation has now an intuitive structure, with the first two terms giving the free energy difference between the surface structure and an equivalent number of bulk alloy atoms. The third term accounts for a possible difference in the surface and bulk stoichiometries and the fourth term for the additional number of O atoms present in the surface geometry. Finally, we note that throughout this text, we use the sign convention that a more negative Gibbs free energy will indicate a more stable state of the system. In the case of a gas phase chemical potential this translates to $\mu_{\rm O_2(gas)}$ approaching $-\infty$ in the limit of an infinitely dilute gas, since adding a particle will then yield an infinite gain in entropy. As a consequence, $\gamma > 0$ indicates the cost of creating the surface between the solid bulk phase and the homogeneous gas phase.

\subsection{Dependence on gas phase and alloy bulk reservoir}

The influence of the surrounding gas phase on the surface structure and composition enters eq. (\ref{general_result}) through the term $N_{\rm O}/2 \; \mu_{\rm O_2(gas)}(T,p_{\rm O_2})$. In increasingly O-rich environments, this term will favor surface structures containing an increasing number of oxygen atoms per surface area. In our approach we assess this effect by computing the surface free energy for a range of oxygen chemical potentials. At sufficiently low $\mu_{\rm O_2(gas)}$ (O-poor conditions), only clean surface structures without oxygen atoms will be stable, which then forms a natural lower bound for the considered range of oxygen chemical potentials. As an upper bound, we use $\mu_{\rm O_2(gas)}({\rm O-rich}) = E_{\rm O_2(gas)}^{\rm total}$, where $E_{\rm O_2(gas)}^{\rm total}$ is the total energy of an isolated O$_2$ molecule (including the zero-point energy). This roughly corresponds to conditions where oxygen would start to condense at the sample at sufficiently low temperatures \cite{reuter02a}. 

At the accuracy level of interest to our study, the oxygen gas phase is well described by ideal gas laws, which then allows to readily relate the oxygen chemical potential to specific temperature and pressure conditions. For this we write \cite{reuter02a}
\begin{eqnarray}
\label{oxygenpT}
&& \mu_{\rm O_2(gas)}(T,p_{\rm O_2}) \;=\; E_{\rm O_2(gas)}^{\rm total} + 2 \Delta \mu_{\rm O}(T,p_{\rm O_2}) \\ \nonumber
&& =\; E_{\rm O_2(gas)}^{\rm total} + 2 \Delta \mu_{\rm O}(T,p^{0}) + k_BT \; {\rm ln} \left(\frac{p_{\rm O_2}}{p^{0}} \right) \quad ,
\end{eqnarray}
where $k_B$ is the Boltzmann constant, and $\Delta \mu_{\rm O}$ contains now all temperature and pressure dependent free energy contributions of the O$_2$ molecule internal degrees of freedom. $\Delta \mu_{\rm O}(T,p_{\rm O_2})$ can be obtained can be obtained within the ideal gas approximation, or alternatively, one can use tabulated enthalpy and entropy values at standard pressure $p^{0} = 1$\,atm to determine $\Delta \mu_{\rm O}(T,p^{0})$ \cite{JANAF}. For an oxygen gas phase, both approaches yield virtually identical results in the range of temperatures and pressures of interest to our study \cite{reuter02a,mcquarrie76}.

Apart from the dependence on the gas phase chemical potential, eq. (\ref{general_result}) shows that the surface free energy depends also critically on the difference of the chemical potentials of the metal atoms $\Delta \mu_{\rm Ag-Pd}$. This variable reflects the change in bulk chemical potential due to changes in $x_{\rm Ag(bulk)}$, as can be seen by evaluating the derivative of eq. (\ref{mu_bulk})
\begin{eqnarray}
\frac{\partial \mu_{\rm bulk}}{\partial x_{\rm Ag(bulk)}} &=&
\mu_{\rm Ag} - \mu_{\rm Pd} + \\ \nonumber
&& \!\!\!\!\!\!\!\!\!\! \left[ x_{\rm Ag(bulk)} 
\frac{\partial \mu_{\rm Ag}}{\partial x_{\rm Ag(bulk)}} + (1-x_{\rm Ag}) 
\frac{\partial \mu_{\rm Pd}}{\partial x_{\rm Ag(bulk)}} \right]  \\ \nonumber
&=& \mu_{\rm Ag} - \mu_{\rm Pd} \;=\; \Delta \mu_{\rm Ag-Pd} \quad ,  
\label{deltamu}
\end{eqnarray} 
where the term in brackets is equal to zero by the Gibbs-Duhem equation \cite{mcquarrie76}. The value of $\Delta \mu_{\rm Ag-Pd}$ will therefore depend sensitively on the bulk reservoir structure and composition, e.g. whether the bulk reservoir is ordered or random, or whether there is a slight excess of one type of atom. This will vary from sample to sample, and can have a significant impact on the surface energetics, even for samples with equal nominal bulk composition \cite{blum02}. 

In our application below, we will demonstrate this crucial dependence on the detailed bulk reservoir structure and composition by considering the case of just a small amount of non-stoichiometries (in form of anti-sites) in the ordered Ag$_3$Pd bulk alloy. Rather than focusing on specific data for a specific sample, the aim is thereby to discuss the general effects that such bulk anti-sites can have on adsorption-induced segregation. Instead of explicitly computing $\Delta \mu_{\rm Ag-Pd}$ e.g. by means of cluster expansion techniques \cite{muller03,fahnle05,sanchez84,fontaine94,zunger94,walle02}, we therefore proceed by roughly estimating the range that $\Delta \mu_{\rm Ag-Pd}$ can span, and discuss the results below for the obtained two bounds. For the estimate we use as limits the phase separation of Ag, when the bulk reservoir is rich in Ag (Ag-rich limit), and the phase separation of Pd, when the bulk reservoir is rich in Pd (Pd-rich limit). Phase separation of Ag will occur, when $\mu_{\rm Ag} > \mu_{\rm Ag,fcc}$, where $\mu_{\rm Ag,fcc}$ is the chemical potential of Ag atoms in the fcc bulk structure of the parent metal. Equivalently, phase separation of Pd will occur, when $\mu_{\rm Pd} > \mu_{\rm Pd,fcc}$, where $\mu_{\rm Pd,fcc}$ is the chemical potential of Pd atoms in the fcc bulk structure of the parent metal. Using eq. (\ref{mu_bulk}) and approximating $\mu_{\rm bulk}$ with its value for the perfectly stoichiometric alloy sample $\mu_{\rm bulk(stoich)}$, we therefore arrive at
\begin{equation}
\Delta \mu_{\rm Ag-Pd(Ag-rich)} \leq
\Delta \mu_{\rm Ag-Pd} \leq
\Delta \mu_{\rm Ag-Pd(Pd-rich)}
\label{mu_range}
\end{equation}
with
\begin{equation}
\Delta \mu_{\rm Ag-Pd(Ag-rich)} \;=\;
\left(\frac{\mu_{\rm Ag,fcc} - \mu_{\rm bulk(stoich)}}{1 - x_{\rm Ag(bulk)}} \right)
\label{Ag-rich}
\end{equation}
and
\begin{equation}
\Delta \mu_{\rm Ag-Pd(Pd-rich)} \;=\;
\left(\frac{\mu_{\rm bulk(stoich)} - \mu_{\rm Pd,fcc}}{1 - x_{\rm Pd(bulk)}} \right)
\label{Pd-rich}
\end{equation}
as a first estimate for the range of $\Delta \mu_{\rm Ag-Pd}$ due to non-stoichiometries in the ordered bulk alloy reservoir.

\subsection{Solid phase Gibbs free energies}

In order to evaluate the surface free energy of a specific configuration with eqs. (\ref{general_result}) and (\ref{mu_range}) we need to compute the solid phase Gibbs free energies, $G_{\rm slab}$, $\mu_{\rm bulk(stoich)}$, $\mu_{\rm Ag,fcc}$ and $\mu_{\rm Pd,fcc}$. For this it is useful to decompose them into several contributing terms, namely
\begin{equation}
G \;=\; E^{\rm total} + F^{\rm vib} + TS^{\rm conf.} + pV \quad ,
\end{equation}
where $E^{\rm total}$ is the total energy (excluding the zero point energy), $F^{\rm vib}$ the vibrational free energy (including the zero point energy), and $S^{\rm conf.}$ the configurational entropy. A crucial aspect that governs our analysis of all of these terms is that the quantity of interest to us, namely the surface free energy, does not depend on absolute Gibbs free energies. What enters into eq. (\ref{general_result}) is a {\em difference} of Gibbs free energies of slab, bulk alloy and bulk metals. This can allow for quite some degree of cancellation, e.g. due to similar free energy contributions in the three systems or due to similar errors in the computed total energies.

We start our analysis by noting that the $pV$ term is completely negligible for the transition metal alloys discussed here \cite{reuter02a}. As for the vibrational free energy, a proper evaluation would require a systematic computation of all vibrational frequencies at the surface and in the bulk systems \cite{reuter02a}. Here, it can be particularly beneficial to realize that only the differences of the vibrational free energy contributions matter for eq. (\ref{general_result}), i.e. the surface free energy is only affected by the {\em changes} of the vibrational modes. In the application discussed below, we will address the effect of on-surface O adsorption on the segregation profile of a Ag$_3$Pd(111) surface. In this case, the vibrational modes of Ag and Pd atoms in the bulk alloy and bulk metal structures will be rather similar, as will be their vibrational modes at the close-packed surface. From the similarity of the calculated stretch mode of oxygen adsorbed at low coverage at Ag(111) and Pd(111), 50\,meV \cite{li02} and 60\,meV \cite{eichler00} respectively, we furthermore do not expect significant variations in the vibrational free energies of alloy surface structures with different metal composition. What will therefore mostly contribute are the changes of the vibrational modes of oxygen in the gas phase and adsorbed at the surface. To estimate this, we use a simple Einstein model equivalent to the one employed in Ref. \onlinecite{reuter02a} and compute the vibrational free energy contribution when the characteristic frequency is changed from $\sim 200$\,meV (O$_2$ stretch frequency in the gas phase) to $\sim 55$\,meV in the adsorbed state. Allowing for O coverages up to 1\,ML and varying the characteristic frequency in the adsorbed state by $\pm 50$\,\%, already the total contribution to the surface free energy stayed always below 5\,meV/{\AA}$^2$ for temperatures ranging up to 600\,K. What matters for the discussion of O-induced segregation is, however, the difference of this vibrational surface free energy contribution for different surface configurations. For surface configurations with the same O coverage (differing only in the Ag and Pd concentration in the surface layers) this difference will be completely negligible in view of the above cited similarity of the O-metal stretch modes. For surface configurations differing in O coverage, the difference in the surface vibrational free energy will be a bit larger, yet still negligible compared to the total energy terms discussed below which are of the order of tens of meV/{\AA}$^2$. Correspondingly, we will neglect the zero point and vibrational free energy contribution to the surface free energy in this study, but emphasize that this is not a general result, and the effect of vibrations needs to be carefully assessed in each application.

This leaves as remaining term the configurational entropy. A full evaluation of this contribution is computationally involved, since it requires a proper sampling of the huge configuration space spanned by all possible bulk and surface structures. Although modern statistical mechanics methods like Monte Carlo simulations \cite{frenkel02,landau02} are particularly designed to efficiently fulfill this purpose, they still require a prohibitively large number of total energy evaluations to be directly linked with electronic structure theories \cite{reuter05}. A way to circumvent this problem is to map the real system onto a simpler, typically discretized model system, the Hamiltonian of which is sufficiently fast to evaluate. Cluster expansions (also termed lattice-gas hamiltonian approach) \cite{muller03,fahnle05,sanchez84,fontaine94,zunger94,walle02} are a prominent example for such a mapping approach, where the considered system would in practice e.g. be described by a lattice of possible on-surface adsorption sites for the gas phase species \cite{stampfl99,han05}. Here, we will instead concentrate on screening a number of known (or possibly relevant or representative) ordered surface structures by directly comparing which of them turns out to be most stable under which ($T,p_{\rm O_2}$)-conditions, i.e. which of them exhibits the lowest surface free energy. For sufficiently low temperatures, the remaining configurational entropy per surface area in $G_{\rm slab}$ is then only due to a limited number of defects in these ordered surface structures and can be estimated to be below 2 meV/{\AA}$^2$ for any $T < 600$\,K \cite{reuter03}. We will restrict the discussion of the application to Ag$_3$Pd(111) below to the temperature range well below this limit, where then also the consideration of an ordered Ag$_3$Pd bulk structure with a small amount of anti-sites is justified \cite{muller01}. Focusing on this temperature range we will neglect the configurational entropy contribution to $G_{\rm slab}$, as well as the small configurational entropy of the ordered stoichiometric alloy and metal bulk reservoirs (entering $\mu_{\rm bulk(stoich)}$, $\mu_{\rm Ag,fcc}$ and $\mu_{\rm Pd,fcc}$). The only remaining temperature dependence in our model comes then from the entropy of the oxygen gas phase (entering $\Delta \mu_{\rm O}(T,p)$), as well as somehow implicitly through the considered bulk alloy chemical potential range, cf. eq. (\ref{mu_range}). This will not affect the conclusions drawn below, but we note that the additional limitation that comes with the then resulting direct screening approach is that its reliability is restricted to the number of considered configurations, or in other words that only the stability of those structures plugged in can be compared. Its predictive power extends therefore only to those structures that are directly considered, i.e. the existence of unanticipated surface geometries or stoichiometries cannot be predicted. With this in mind, the here employed direct screening approach to first-principles atomistic thermodynamics can still be a particularly valuable tool, since it allows, for example, to rapidly compare the stability of different structural models without the need to have them mapped onto a common lattice. In this sense, we view the present approach as complementary to a cluster expansion study, since it provides first insight into the possible effects of the surrounding gas phase by screening a wide range of possible structures. If more detailed insight is required (e.g. the exact surface concentration profile), this can then selectively be refined by a cluster expansion study, in particular when phase coexistence or order-disorder phenomena at elevated temperatures are of interest.

In the present work, we therefore approximate the difference of Gibbs free energies entering eq. (\ref{general_result}) by the corresponding difference of the leading total energy terms, and our final working equation reads
\begin{eqnarray}
\label{work_eq}
\lefteqn{\gamma(T,p_{\rm O_2},N_{\rm O},N_{\rm slab},x_{\rm Ag},x_{\rm Pd}) \;\approx} \\ \nonumber
&& \frac{1}{2 A_{\rm surf}} \; \bigg[ E^{\rm total}_{\rm slab}(N_{\rm O},N_{\rm slab},x_{\rm Ag},x_{\rm Pd}) \; - \\ \nonumber &&  N_{\rm slab} E^{\rm total}_{\rm bulk(stoich)} -
N_{\rm slab} \Delta \mu_{\rm Ag-Pd} (x_{\rm Ag} - x_{\rm Ag(bulk)}) \; - \\ \nonumber
&& -\; \frac{N_{\rm O}}{2} E^{\rm total}_{\rm O_2(gas)}
- N_{\rm O} \Delta \mu_{\rm O}(T,p_{\rm O_2}) \bigg] \quad .
\end{eqnarray}
What is then required in practice to evaluate the surface energy of a given surface geometry of Ag$_3$Pd(111) in a given oxygen gas phase characterized by $\Delta \mu_{\rm O}(T,p_{\rm O_2})$ are total energy calculations of $E^{\rm total}_{\rm slab}(N_{\rm O},N_{\rm slab},x_{\rm Ag},x_{\rm Pd})$ of the surface (in the supercell geometry), of $E^{\rm total}_{\rm bulk(stoich)}$ of the stoichiometric Ag$_3$Pd alloy bulk, and of $E^{\rm total}_{\rm O_2(gas)}$ of an isolated O$_2$ molecule. The considered range for the bulk alloy chemical potential difference $\Delta \mu^{\rm total}_{\rm Ag-Pd}$ is given through eq. (\ref{mu_range}), for which (with the equivalent approximations) total energy calculations of the parent Ag and Pd bulk metals ($E^{\rm total}_{\rm Ag, fcc}$, $E^{\rm total}_{\rm Pd, fcc}$) are additionally required. If we define the formation energy for the bulk alloy as
\begin{eqnarray}
\nonumber
\lefteqn{\Delta E_f({\rm Ag}_{n_{\rm Ag(bulk)}}{\rm Pd}_{n_{\rm Pd(bulk)}}) \;=} \\
\nonumber
&& E^{\rm total}_{\rm bulk(stoich)} \;-\; 
x_{\rm Ag(bulk)} E^{\rm total}_{\rm Ag, fcc} \;- \\
&& (1 - x_{\rm Ag(bulk)}) E^{\rm total}_{\rm Pd, fcc} \quad ,
\label{formationeng}
\end{eqnarray}
then this range spanned by $\Delta \mu_{\rm Ag-Pd}$ is given by
\begin{eqnarray}
\label{murange_work}
\nonumber
\lefteqn{\Delta \mu_{\rm Ag-Pd(Ag-rich)} - \Delta \mu_{\rm Ag-Pd(Pd-rich)} \;\approx} \\
&& \frac{\Delta E_f({\rm Ag}_{n_{\rm Ag(bulk)}}{\rm Pd}_{n_{\rm Pd(bulk)}})}{x_{\rm Ag(bulk)} \left( 1 - x_{\rm Ag(bulk)} \right) } \quad .
\end{eqnarray}

\subsection{Total energy calculations}

We compute all total energies within pseudopotential plane wave density-functional theory (DFT) \cite{dacapo}, using the generalized gradient approximation (GGA) \cite{perdew96} for the exchange-correlation functional. The employed Pd ultrasoft pseudopotential (USPP) \cite{vanderbilt90,laasonen93,uspp} has a $4d^95s^1$ electronic configuration and is based on a Koelling-Harmon relativistic all-electron calculation \cite{koelling77}. The cutoff radii were set at 2.5, 2.5, 2.0 bohr for the $s$, $p$ and $d$ channels. The Ag USPP has a $4d^{10}5s^{1}$ electronic configuration, with cutoff radii set at 2.5, 2.5, 2.1 bohr for the $s$, $p$ and $d$ channels. For both elements the $f$ angular momentum channel is used as the local potential, with two $s$, $p$ and $d$ projectors, and a non-linear core correction \cite{louie82} is used with $r_{\rm core} = 1.2$ bohr. The O USPP has a $2s^2 2p^4$ electronic configuration, with cutoff radii set at 1.3 bohr for the $s$ and $p$ channels. The $d$ angular momentum channel is used as the local potential, with two $s$ and $p$ projectors, and a non-linear core correction with $r_{\rm core} = 0.5$ bohr. We found it advantageous to use a less accurate oxygen USPP with a larger partial core radius ($r_{\rm core} = 0.7$ bohr) to pre-relax our geometries at a lower computational cost. Substituting the more accurate O USPP in a final calculation, we found the maximum forces to be very close to or less than the force criteria set in the initial optimization ($< 0.05$\,eV/{\AA}). In several cases where the maximum forces exceeded the optimization criteria, we reoptimized the geometry, but found that the surface energies changed by less than 1\,meV/{\AA}$^2$, which is accepted as a negligible error in the remainder of the calculations. A plane wave cutoff of 350\,eV was employed in all calculations using the less accurate O USPP, while we employed the double grid method \cite{laasonen93} in the calculations using the more accurate O USPP. The higher energy cutoff (700\,eV) with which the density is treated in the latter method increases the computational time moderately, but resulted in smooth convergence behavior \cite{hammer99}.

In order to ensure maximum compatibility between the bulk metal and bulk alloy calculations, we employed the same bulk unit-cell to describe both the L1$_2$ Ag$_3$Pd and the fcc Pd and fcc Ag structures. $(20 \times 20 \times 20)$ Monkhorst-Pack {\bf k}-point grids were then used to optimize the lattice constants to $a_{0}({\rm Ag_3Pd}) = 4.09$\,{\AA}, $a_{0}({\rm Pd}) = 3.93$\,{\AA}, and $a_{0}({\rm Ag}) = 4.09$\,{\AA}. Using these optimized lattice constants, we obtain a formation energy $\Delta E_f({\rm Ag_3Pd})$ as defined in eq. (\ref{formationeng}) of $-$0.21\,eV per formula unit. The (111) alloy surfaces are modeled in a periodic supercell geometry, employing 7-layer inversion-symmetric slabs with two identical surfaces. The outer two layers on each side of the slab were fully relaxed, keeping the inner three layers fixed. A vacuum region of 14\,{\AA} ensures the decoupling of consecutive slabs. The Brillouin zone integrations were performed with $(8 \times 8 \times 1)$ Monkhorst-Pack {\bf k}-point grids for the $(2 \times 2)$ surface unit-cells. For the oxygen adsorption calculations, oxygen was placed on each side of the slabs, maintaining inversion symmetry in all cases. For the computation of $E^{\rm total}_{\rm O_2(gas)}$ we employed a $(12 \times 13 \times 14)$\,bohr cell and obtain an O$_2$ binding energy of $-$5.91\,eV. From test calculations with up to twice as many {\bf k}-points and an increased plane wave cutoff of 500\,eV we estimate the numerical accuracy of the reported surface energies to be within 2\,meV/{\AA}$^2$, which does not affect any of the physical conclusions drawn.

\section{Results}

\subsection{O-induced Pd segregation at Ag$_3$Pd(111)}

\begin{figure}
\scalebox{0.6}{\includegraphics{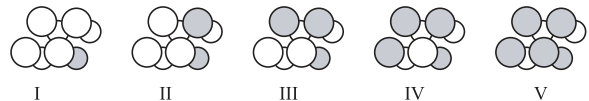}}
\caption{Top views of the atomic geometry of five ($2\times 2$) surface unit-cell arrangements considered in this study. The shown configurations have a bulk like second layer composition, but a first layer composition that varies from pure Ag (I) to pure Pd (V). White colored spheres correspond to Ag atoms, whereas grey colored spheres correspond to Pd atoms. The larger spheres correspond to the first layer, and the smaller ones correspond to the second layer.}
\label{fig1}
\end{figure}

We illustrate our approach by addressing the O-induced segregation profile at a Ag$_3$Pd(111) surface. At the nominal stoichiometry each (111) layer in the assumed L1$_2$ bulk crystal structure \cite{muller01} contains 3 Ag atoms and 1 Pd atom in a $(2 \times 2)$ arrangement. We account for a possible segregation at the surface by considering $(2 \times 2)$ surface unit-cell atomic configurations in which the composition in the topmost two layers is systematically varied from pure Ag to pure Pd. This leads to 25 possible substrate configurations, five of which are shown in Fig. \ref{fig1}. In addition to these clean surface structures, oxygen adsorption was then examined by placing from one up to four O atoms per surface unit-cell into the on-surface fcc sites. This leads to a pool of 473 inequivalent surface structures (including some larger unit cells), in which the oxygen coverage varies from 0 to 1 monolayer (ML) in $1/4$ ML steps in the $(2 \times 2)$ surface unit-cells.

\begin{figure}
\scalebox{0.9}{\includegraphics{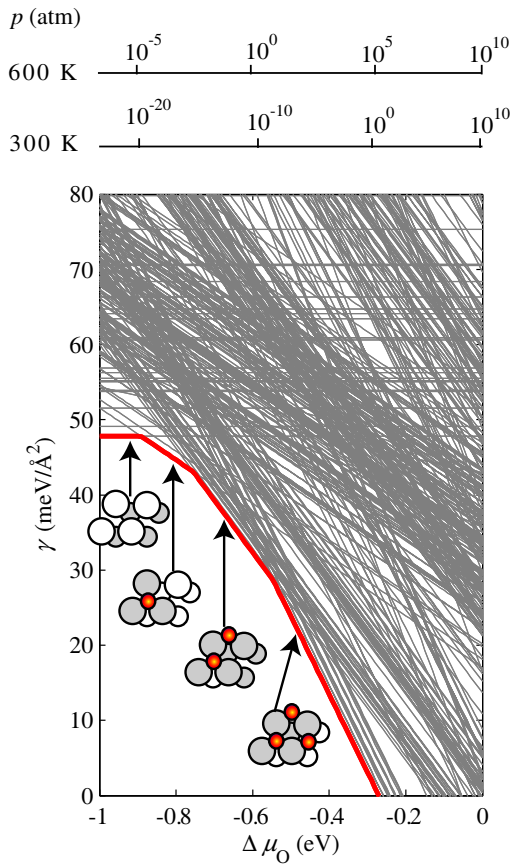}}
\caption{(Color online) Surface free energy of Ag$_3$Pd(111) in equilibrium with a Pd-rich Ag$_3$Pd bulk reservoir, cf. eq. (\ref{Pd-rich}), as a function of oxygen chemical potential. Each line corresponds to one of the tested surface configurations, and only the few configurations that result as most stable for a range of oxygen chemical potential are drawn as dark (red) lines. Additionally shown as insets are top views of the most stable surface configurations in the same form as in Fig. \ref{fig1}, with adsorbed O atoms shown as dark small circles, Ag atoms as white circles and Pd atoms as grey circles. The dependence on the oxygen chemical potential is translated into pressure scales using eq. (\ref{oxygenpT}) for $T=300$\,K and $T=600$\,K (upper $x$-axes).\label{fig2}}
\end{figure}

\begin{figure}
\scalebox{0.9}{\includegraphics{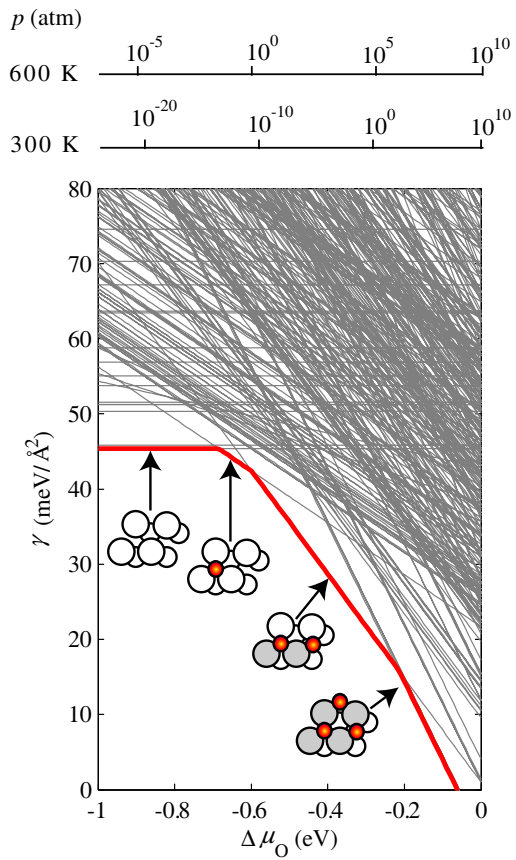}}
\caption{(Color online) Surface free energy Ag$_3$Pd(111) in equilibrium with a Ag-rich Ag$_3$Pd bulk reservoir, cf. eq. (\ref{Ag-rich}), as a function of oxygen chemical potential. Each line corresponds to one of the tested surface configurations, and only the few configurations that result as most stable for a range of oxygen chemical potential are drawn as colored lines. Additionally shown as insets are top views of the most stable surface configurations in the same form as in Fig. \ref{fig1}, with adsorbed O atoms shown as dark small circles, Ag atoms as white circles and Pd atoms as grey circles. The dependence on the oxygen chemical potential is translated into pressure scales using eq. (\ref{oxygenpT}) for $T=300$\,K and $T=600$\,K (upper $x$-axes).\label{fig3}}
\end{figure}

Evaluating eq. (\ref{work_eq}) for all configurations we arrive at the surface free energy plots shown in Figs. \ref{fig2} and \ref{fig3}. Figure \ref{fig2} corresponds to the limiting case of a Pd-rich Ag$_3$Pd bulk reservoir and Fig. \ref{fig3} to the opposite limit of a Ag-rich Ag$_3$Pd bulk reservoir as discussed in Section IIB, cf. eq. (\ref{mu_range}). Within our thermodynamic theory, at each oxygen chemical potential the surface configuration yielding the lowest surface free energy (i.e. the lowest line in the figures) results as ``most stable''. Focusing first on the Pd-rich situation, we observe different stable configurations in different oxygen chemical potential ranges. At the lowest chemical potentials considered, the lowest surface free energies are naturally obtained for the clean surface structures with varying chemical composition in the topmost two layers. Among those, the ``most stable'' surface configuration under these O-poor environmental conditions (i.e. the lowest horizontal line in Fig. \ref{fig2}) corresponds to a Ag-terminated alloy, with 100\,\% Ag in the first layer and 100\,\% Pd in the second layer. As expected\cite{reniers95}, we therefore obtain segregation of the more noble metal to the surface under gas phase conditions that at room temperature are representative of ultra-high vacuum experiments, cf. the pressure scales in Fig. \ref{fig2}.

Due to the last term in eq. (\ref{work_eq}), increasingly O-rich surface structures get more favorable with increasing $\Delta \mu_{\rm O}$. The higher the O concentration ($N_{\rm O}/A_{\rm surf}$) at the surface, the steeper the slope of the lines in Figs. \ref{fig2} and \ref{fig3}. In the case of the Pd-rich bulk limit, a first oxygen containing surface structure gets more stable than the clean Ag-segregated surface above $\Delta \mu_{\rm O} = -0.89$\,eV. This configuration corresponds to an oxygen coverage of a 1/4\,ML and contains now already 75\% Pd atoms in the first layer. At even higher O chemical potentials further surface structures with higher oxygen coverages get stabilized, each time then containing a first layer composition of only Pd atoms. Induced by the oxygen adsorption is thus a segregation of the more reactive metal to the surface, which effectively reverses the segregation profile corresponding to room temperature ultra-high vacuum gas phase conditions. 

While this intuitive trend comes out clearly from our theory accounting for the effect of the environment, the exact surface concentrations as reflected by the obtained ``most stable'' structures have to be considered with care. Our approach considers a finite set of ordered surface structures and neglects the configurational entropy contribution to the solid phase Gibbs free energies as discussed in Section IIC. The closeness of the lowest energy lines in Fig. \ref{fig2} reflects a number of structures that are just a little bit less stable at a given $\Delta \mu_{\rm O}$ than the highlighted ``most stable'' structure. This indicates already that a more exhaustive search of the huge configuration space e.g. by means of a cluster expansion technique would likely reveal new ground state structures with compositions that are intermediate to those considered here and that exhibit a certain degree of disorder at finite temperatures.

While a quantitative determination of the segregation profile is therefore outside the scope of the present work, the direct comparison of Figs. \ref{fig2} and \ref{fig3} nicely illustrates the entangled dependencies of the surface composition on both the bulk alloy {\em and} the gas phase reservoir. The results for the Ag-rich bulk alloy reservoir summarized in Fig. \ref{fig3} also indicate an O-induced Pd segregation at increasing $\Delta \mu_{\rm O}$. One major difference to the afore discussed Pd-rich bulk alloy reservoir is that the stabilization of these O-containing Pd-enriched surface structures occurs at higher oxygen chemical potentials. This obviously reflects the higher cost at which the additional Pd surface atoms have to be taken out of the Ag-rich alloy bulk reservoir, and correspondingly a higher driving force from the gas phase is required to establish the same Pd content in the surface layers as compared to the Pd-rich bulk alloy reservoir. A closer look at Figs. \ref{fig2} and \ref{fig3}, and equivalently inspection of eq. (\ref{work_eq}), reveals, however, that the effect of the different bulk alloy reservoir is not just a mere shift of all surface free energy curves to higher $\Delta \mu_{\rm O}$ by an amount that corresponds to the range of $\Delta \mu_{\rm Ag-Pd}$ between our two considered bulk alloy reservoir limits, cf. eq. (\ref{murange_work}). Within our computational setup this range amounts to 0.28\,eV, and this is indeed roughly the order of magnitude by which the lines in Fig. \ref{fig3} are shifted compared to those in Fig. \ref{fig2}. Still, due to the prefactor $(x_{\rm Ag} - x_{\rm Ag(bulk)})$ of the $\Delta \mu_{\rm Ag-Pd}$-term in eq. (\ref{work_eq}) surface energies of different surface configurations are affected differently by the change in the bulk reservoir. Accordingly, we observe also a different sequence of ``most stable'' structures in Fig. \ref{fig3} compared to Fig. \ref{fig2}, which underscores the need to explicitly account for both the effect of the environment and the bulk alloy structure to describe the structure and composition of alloy surfaces in realistic applications.

Another intricacy that is nicely illustrated by our data is the effect of compositional changes in the deeper substrate layers. As apparent from Figs. \ref{fig2} and \ref{fig3} the first-layer Pd-segregation induced in more oxygen-rich environments is accompanied by substantial changes of the second layer composition. These changes have quite some impact on the stability of the various surface structures as is revealed when restricting our structure data base to those structures with nominal Ag$_3$Pd second layer stoichiometry. When evaluating eq. (\ref{work_eq}) for this subset of surface structures we obtain as 
``most stable'' clean surface for the case of a Pd-rich bulk reservoir a structure that contains only 75\% Ag in the first layer (in contrast to the 100\% Ag termination when also considering non-stoichiometries in the second substrate layer). While the remaining sequence of ``most stable'' structures with increasing Pd content in the first layer is similar to the one shown in Figs. \ref{fig2} and \ref{fig3} respectively, the oxygen chemical potentials at which the transitions between the different structures occur are shifted by sometimes more than 0.1\,eV compared to the values exhibited in Figs. \ref{fig2} and \ref{fig3}. As indicated by the drawn pressure scales in the figures this can correspond to several orders of magnitude in pressure in the temperature range discussed here, from which we conclude that a quantitative modeling of the Ag$_3$Pd(111) surface requires at least the consideration of non-stoichiometries in the topmost two substrate layers.

\subsection{Contributing factors to adsorbate-induced segregation}

The results presented in the previous section clearly show that the adsorption-induced segregation profile at the surface results from the compromise between the tendency to lower the surface free energy by segregating the more noble metal species to the surface, and the tendency to lower the surface free energy by achieving stronger adsorbate bonding at surfaces enriched with the more reactive metal species. In order to further qualify these two competing trends we define as a fundamental quantity describing the adsorbate bonding the average oxygen binding energy $E_{\rm b}(N_{\rm O},x_{\rm Ag},x_{\rm Pd})$ at a given surface configuration containing $N_{\rm O}$ adsorbed O atoms per surface unit-cell and with a detailed surface composition in the surface layers characterized by mole fractions of $x_{\rm Ag}$ Ag atoms and $x_{\rm Pd}$ Pd atoms,
\begin{eqnarray}
\nonumber
\lefteqn{E_{\rm b}(N_{\rm O},x_{\rm Ag},x_{\rm Pd}) \;=\;} \\
\nonumber
&& \frac{1}{2N_{\rm O}} \left[ E^{\rm total}_{\rm slab}(N_{\rm O},N_{\rm slab},x_{\rm Ag},x_{\rm Pd})
\;- \right.  \\
&& \left. E^{\rm total}_{\rm slab}(0,N_{\rm slab},x_{\rm Ag},x_{\rm Pd}) \;-\; \frac{N_{\rm O}}{2} E^{\rm total}_{\rm O_2(gas)} \right] \quad .
\end{eqnarray}
As before this definition holds directly for symmetric slabs with adsorption on both sides, and with the chosen sign convention and zero reference in form of the gas phase O$_2$ molecule an average binding energy $E_{\rm b} < 0$ indicates that dissociative adsorption is exothermic.

Similarly we define as a quantity describing the surface segregation the segregation energy $E_{\rm seg}(x_{\rm Ag},x_{\rm Pd})$ as the cost to change the reference clean alloy surface with nominal bulk stoichiometric composition into an alternative clean surface configuration with concentrations of $x_{\rm Ag}$ Ag atoms and $x_{\rm Pd}$ Pd atoms in the topmost layers
(either by rearrangement of the surface atoms including atoms in sub-surface layers, or by segregation of atoms to or from the bulk reservoir),
\begin{eqnarray}
\label{segeng}
\nonumber
\lefteqn{E_{\rm seg}(x_{\rm Ag},x_{\rm Pd}) \;=\; \frac{1}{2} \left[ E^{\rm total}_{\rm slab}(0,N_{\rm slab},x_{\rm Ag},x_{\rm Pd}) \;- \right. } \\
\nonumber
&& E^{\rm total}_{\rm slab}(0,N_{\rm slab},x_{\rm Ag(bulk)},x_{\rm Pd(bulk)}) \;- \\
&& \left . N_{\rm slab} \Delta E^{\rm total}_{\rm Ag-Pd} (x_{\rm Ag} - x_{\rm Ag(bulk)}) \right] \quad .
\end{eqnarray}
The factor $1/2$ accounts as before for the fact that we are considering symmetric slabs where the segregation occurs equally at both surfaces, and within our sign convention a segregation energy $E_{\rm seg} < 0$ indicates that segregation is exothermic.

With these definitions, we can now recast our working equation for the surface free energy, eq. (\ref{work_eq}), into an intuitive structure,
\begin{eqnarray}
\label{contributions}
\lefteqn{\gamma(T,p_{\rm O_2},N_{\rm O},N_{\rm slab},x_{\rm Ag},x_{\rm Pd}) \;=} \\
\nonumber
&& \gamma_{\rm bulk-term.} \;+\; \frac{E_{\rm seg}(x_{\rm Ag},x_{\rm Pd})}{A_{\rm surf}} \;+ \\
\nonumber
&& N_{\rm O} \frac{E_{\rm b}(N_{\rm O},x_{\rm Ag},x_{\rm Pd})}{A_{\rm surf}} \;-
\frac{N_{\rm O}}{2A_{\rm surf}} \Delta \mu_{\rm O}(T,p_{\rm O_2}) \quad ,
\end{eqnarray}
where $\gamma_{\rm bulk-term.}$ is the surface free energy of the reference clean alloy surface with nominal bulk stoichiometric composition. This form of the equation shows clearly that the surface free energy can either be lowered by 
an energetically favorable segregation of metal atoms ($E_{\rm seg} < 0$), or by exothermic adsorption of $N_{\rm O}$ oxygen atoms ($E_{\rm b} < 0$), where increasing oxygen chemical potentials in the gas phase further favor configurations with an increasing number of oxygen atoms at the surface. Having disentangled these various contributing factors, we now proceed to discuss our detailed results for each one separately. This enables us to carve out in detail the complexity behind each of them, thereby bringing us into the position to scrutinize various approximate treatments that have been suggested in the literature.

\subsubsection{Segregation energy}

\begin{figure}
\scalebox{0.9}{\includegraphics{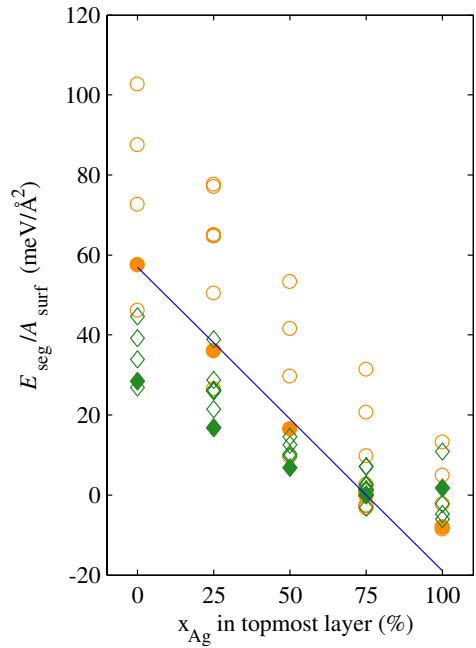}}
\caption{(Color online) Segregation energies per surface area, $E_{\rm seg}/A_{\rm surf}$, for all clean surface configurations considered and as a function of the Ag concentration in the topmost layer. (Yellow) circles show the segregation energies for the Ag-rich bulk reservoir and (green) rhombes the segregation energies for the Pd-rich bulk reservoir. The different data points at the same topmost layer Ag concentration represent the variations due to differing Ag concentration in the second layer, and the solid circles viz. rhombes correspond to the surface configurations with nominal bulk composition in the second layer. Additionally shown by the blue line is the segregation energy estimate obtained by considering the segregation of diluted Ag impurities in Pd and diluted Pd impurities in Ag (see text).\label{fig4}}
\end{figure}

Figure \ref{fig4} shows the segregation energies as defined in eq. (\ref{segeng}) for the 25 clean Ag$_3$Pd(111) surface configurations computed. As expected, the segregation energy decreases overall largely with increasing Ag content in the topmost layer, which is the driving force behind the stabilization of the Ag-terminated clean alloy surface structures at the lowest oxygen chemical potentials in Figs. \ref{fig2} and \ref{fig3}. In detail, this variation of the segregation energy is, however, not simple, and depends much on the bulk alloy reservoir and the second layer composition. For the 100\% Ag-terminated surfaces, the lowest segregation energy is for example obtained for a 0\% Pd concentration in the second layer in the case of the Ag-rich bulk reservoir, but for a 100\% Pd concentration in the second layer in the case of the Pd-rich bulk reservoir. 

This complexity needs to be captured in the modeling to obtain quantitatively accurate results for the segregation profile, and thus dictates an explicit consideration of the structure and composition of both alloy bulk and surface. A prevalent alternative approach is instead to obtain estimates for the segregation energy by computing the energy cost to segregate an impurity atom in a metal host to the surface \cite{ruban99a,ruban99b}. In the present case, this would correspond to computing a diluted Ag impurity in the bulk or at the (111) surface of fcc Pd, or a diluted Pd impurity in the bulk or at the (111) surface of fcc Ag. Within our computational setup, we obtain such impurity segregation energy estimates by calculating the total energy difference when once placing a Ag atom in the central layer and once in the topmost layer of a seven layer $(2 \times 2)$ Pd(111) slab (at the optimized Pd lattice constant), as well as respectively doing the same for a Pd atom and a Ag(111) slab (at the optimized Ag lattice constant). The values obtained are $-$0.31\,eV for the case of a Ag atom in Pd, and +0.24\,eV for the case of a Pd atom in Ag, which compare well to the equivalent numbers computed by Ruban and coworkers\cite{ruban99b} using the local-density approximation as exchange-correlation functional (-0.26\,eV and +0.28\,eV, respectively). 

Using these numbers the total segregation energy of a given surface configuration is then simply obtained by appropriately summing the impurity segregation costs. For a 100\% Pd-terminated surface this corresponds e.g. to moving three Ag atoms per $(2 \times 2)$ surface unit-cell to the bulk and moving three Pd atoms per $(2 \times 2)$ surface unit-cell from the bulk ($- 3 \times (-0.31$\,eV) + $3 \times 0.24$\,eV = 1.65\,eV). Normalized to the area of the $(2 \times 2)$ Ag$_3$Pd(111) surface unit-cell, this yields $E_{\rm seg}/A_{\rm surf} = 57$\,meV/{\AA}$^2$ in this simple model. The blue line in Fig. \ref{fig4} shows the corresponding data for such variations in the topmost layer Ag concentration, which reproduces the overall trend of the calculated DFT data surprisingly well (in particular when comparing to the filled symbols representing the DFT data where also the second layer composition has been kept at the bulk stoichiometry). By construction, the model can however not account for the dependence on the bulk alloy reservoir and leads thus to quantitative errors of the order of several 10\,meV/{\AA}$^2$ in the $E_{\rm seg}/A_{\rm surf}$ contribution to the surface free energy. This is quite sizable considering the very magnitude of $E_{\rm seg}/A_{\rm surf}$ and $\gamma$ reflected in Figs. \ref{fig2}, \ref{fig3} and \ref{fig4}, and underscores the necessity for a more refined treatment of the segregation energy contribution which does account for the detailed alloy bulk and surface structure.

\subsubsection{Oxygen binding energy}

\begin{figure}
\scalebox{0.9}{\includegraphics{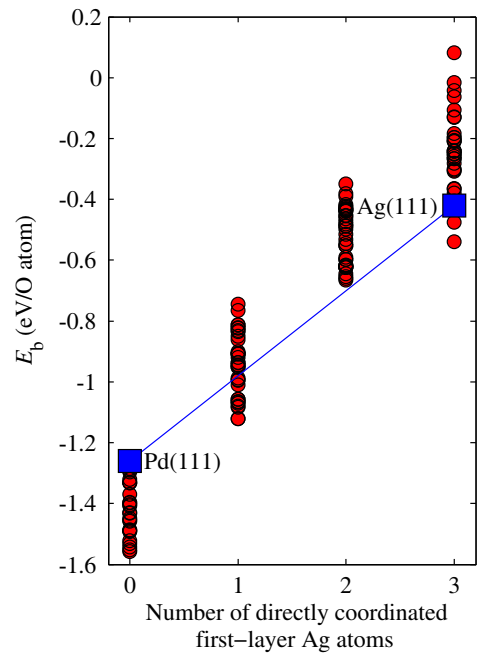}}
\caption{(Color online) Average binding energy $E_{\rm b}$ per O atom as a function of the number of directly coordinated first-layer Ag atoms at a coverage of $\theta =0.25$\,ML, i.e. with one O atom per $(2 \times 2)$ surface unit-cell. For the considered adsorption into fcc hollow sites at the close-packed (111) surface, this number of directly coordinated Ag atoms can range from zero to three. Additionally shown as (blue) squares is the average binding energy at this coverage at the fcc sites of a Pd(111) surface (corresponding to zero Ag atom coordination) and of a Ag(111) surface (corresponding to three Ag atom coordination). The (blue) line is a guide to the eye, representing a simple linear average between the binding energy at these two parent metal surfaces.
\label{fig5}}
\end{figure}

A similar complexity as for the segregation energy is observed, when we turn to the third term in eq. (\ref{contributions}) and therewith to the term that is directly responsible for the oxygen-induced segregation. Starting the discussion with the lowest O coverage that can be treated within our $(2 \times 2)$ surface unit-cells, i.e. 0.25\,ML with 1 O atom per surface unit-cell, Fig. \ref{fig5} shows the variation of the average binding energy with the number of first-layer Ag atoms to which the adsorbed O atom is directly coordinated. As expected, less exothermic binding energies are obtained with an increasing number of Ag atoms immediately involved in the oxygen bonding. This is in line with the known stronger bond strength of oxygen at the more reactive Pd(111) surface compared to the more noble Ag(111) surface. However, the situation at the alloy surface is not a mere linear composition weighted average between these two limiting cases as also shown in Fig. \ref{fig5}. In fact, we find average binding energies at Pd-rich Ag$_3$Pd(111) surface configurations that are even stronger than on Pd(111), and average binding energies at Ag-rich Ag$_3$Pd(111) surface configurations that are even weaker than on Ag(111). Judging from the different, known binding properties at the parent metal surfaces would therefore significantly underestimate the effect of oxygen-induced Pd surface segregation at this alloy surface, i.e. also in this case a more simplified treatment that does not explicitly account for the alloy surface structure would not be appropriate.

The range of average binding energies apparent in Fig. \ref{fig5} for the same number of directly coordinated first-layer Ag atoms is primarily due to the differing second layer compositions of the various surface configurations behind the data points. Interestingly, this effect of a varying number of Pd atoms in the second layer is different for different numbers of Pd atoms in the first layer. For the configurations with Pd-rich first layer compositions, increasing the number of Pd atoms in the second layer increases the average O binding energy, whereas for Ag-rich first layer compositions incorporation of Pd into the second layer substantially reduces the average O binding energy. We find a similar effect of the second layer composition also at all other computed O coverages up to 1\,ML, each time leading to variations of $E_{\rm b}$ of the order of $\sim 0.3-0.6$\,eV per O atom. For the most O-rich surface structures this amounts to variations of the term $N_{\rm O} E_{\rm b}/A_{\rm surf}$ in eq. (\ref{contributions}) of up to $\sim 80$\,meV/{\AA}$^2$, which is then of comparable order, if not larger than the corresponding variation of the $E_{\rm seg}/A_{\rm surf}$ term with non-stoichiometries in the second layer. The non-trivial interplay between these two equally-sized contributions to the surface free energy is thus what dictates at least the consideration of non-stoichiometries in the topmost two layers when aiming at a quantitative description of the adsorbate-induced segregation at this Ag$_3$Pd(111) surface. 

From our data, we also arrive at a first estimate of the impact that the third layer atoms have on the surface adsorption properties. For surface configurations with pure first and second layer compositions, i.e. either 100\,\% Ag or 100\,\% Pd in either of the two layers, the four fcc adsorption sites in our $(2 \times 2)$ surface unit-cells are completely equivalent, except for variations in the type of atoms in the third bulk-stoichiometric layer (i.e. with three Ag atoms and one Pd atom per $(2 \times 2)$ cell). In these cases, we obtained variations of up to 0.1\,eV per O atom in $E_{\rm b}$, depending on whether the oxygen atom was located directly over a Pd atom in the third layer or not. Compared to the average binding energy variations due to differing compositions in the topmost two layers, these are still noticeable, but already much smaller variations. On this basis, we would therefore conclude that a consideration of non-stoichiometries in the topmost two layers should capture the essential physics of oxygen-induced segregation at Ag$_3$Pd(111).

\begin{figure}
\scalebox{0.9}{\includegraphics{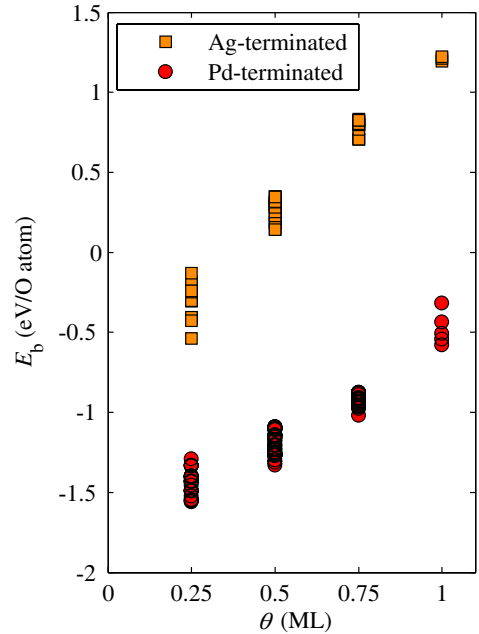}}
\caption{(Color online) Average binding energy $E_{\rm b}$ per O atom as a function of the oxygen coverage, ranging from 0.25\,ML to 1\,ML (corresponding to one to four O atoms per $(2 \times 2)$ surface unit-cell). Shown is only the data for surface configurations with a 100\,\% Ag concentration in the topmost layer (yellow squares) or with a 100\,\% Pd concentration in the topmost layer (red circles). The different data points at the same coverage correspond to surface configurations with varying second layer composition. The full Ag second layer gives the lower energies and the full Pd second layer the higher energies.
\label{fig6}}
\end{figure}

Apart from the dependence on the surface composition the average O binding energy also displays a strong variation as a function of coverage. Figure \ref{fig6} illustrates this for the subset of computed configurations with either a 100\,\% Ag or a 100\,\% Pd concentration in the topmost layer. In both cases, we observe a strong reduction in bond strength with increasing number of oxygen adsorbates at the surface. For the Ag-terminated surfaces, this has the consequence that dissociative adsorption becomes endothermic at all configurations with coverages $\theta > 0.25$\,ML, which is very similar to the findings at the pure Ag(111) surface \cite{todorova02}. Within the scatter in the data shown in Fig. \ref{fig6}, which is due to the varying second layer composition of the surface configurations displayed, the decrease of the average binding energy with coverage is roughly linear for both Ag and Pd terminated surfaces. Again, this is similar to the bonding at the parent metal surfaces Ag(111) and Pd(111) \cite{todorova02}. The amount, by which $E_{\rm b}$ decreases with coverage is, however, not the same for the two differently terminated surfaces. Whereas for the Pd-terminated surfaces the average binding energy decreases by about 1\,eV in the computed coverage range $0.25 \le \theta \le 1$\,ML, it decreases for the Ag-terminated surfaces by about 1.5\,eV. In other words, the anyway stronger bonding at Pd-enriched surfaces at low O coverages becomes even more favored at increasing O coverage in comparison to Ag-enriched surfaces with the same oxygen load. This will further enhance the tendency towards Pd surface segregation at increasing O chemical potentials, and shows that a restricted information about low coverage bonding properties will also not be sufficient to correctly model the segregation profile in a reactive environment like the presently studied oxygen atmospheres.

\subsubsection{Gas phase chemical potential}

Even though the average binding energies per O atom decrease with increasing oxygen coverage, the total contribution of $N_{\rm O}$ times the smaller $E_{\rm b}$ in eq. (\ref{contributions}) may still lead to lowest surface free energies for surface configurations with higher oxygen load. At increasing oxygen chemical potentials such structures are further favored by the last term in eq. (\ref{contributions}), i.e. the increasing abundance of oxygen in the gas phase helps to stabilize O-rich surface structures. Within our calculated set of surface configurations, this leads to the sequence of ``most stable'' structures in Figs. \ref{fig2} and \ref{fig3}, with increasing concentrations of adsorbed O atoms at higher values of $\Delta \mu_{\rm O}$. At sufficiently high oxygen chemical potentials also more O-rich surface configurations beyond the simple on-surface adlayers considered in the present study may get stabilized. Apart from ultra-thin so-called surface oxide films, which have been characterized for both parent metal surfaces \cite{todorova03,schnadt06}, this will ultimately be thicker bulk-like oxide films, and due to their higher stability most likely palladium oxides.

Assuming that what forms at the surface is simply the most stable palladium bulk oxide, PdO, we can estimate the oxygen chemical potential at which such a bulk oxidation occurs from the stability condition \cite{reuter04}
\begin{equation}
\mu_{\rm PdO} \;<\; \mu_{\rm Pd} + \mu_{\rm O} \quad .
\end{equation}
Within the same approximations as applied in Section IIC, we then arrive in the limit of the Pd-rich bulk reservoir ($\mu_{\rm Pd} = \mu_{\rm Pd(metal)}$) at
\begin{equation}
\Delta \mu_{\rm O}^{\rm bulk oxid., Pd-rich} \;>\; \Delta E_f({\rm PdO}) \quad , 
\end{equation}
where the formation energy of bulk PdO is defined as 
\begin{equation}
\Delta E_f({\rm PdO}) \;=\; E^{\rm total}_{\rm PdO, bulk} \;-\; E^{\rm total}_{\rm Pd,fcc} \;-\; 1/2 E^{\rm total}_{\rm O_2(gas)} \quad .
\end{equation}
Similarly, one obtains in the limit of the Ag-rich bulk reservoir ($\mu_{\rm Ag} = \mu_{\rm Ag(metal)}$)
\begin{equation}
\Delta \mu_{\rm O}^{\rm bulk oxid., Ag-rich} \;>\; \Delta E_f({\rm PdO}) \;-\; \Delta E_f({\rm Ag_3Pd}) \quad , 
\end{equation}
with the alloy formation energy as defined in eq. (\ref{formationeng}).

Within our computational setup, we determine a value of $E_f({\rm PdO}) = -0.86$\,eV per formula unit, so that the chemical potentials corresponding to the onset of bulk oxide formation are $\Delta \mu_{\rm O}^{\rm bulk oxid., Pd-rich} = -0.86$\,eV and $\Delta \mu_{\rm O}^{\rm bulk oxid., Ag-rich} = -0.65$\,eV. In both cases, the onset of bulk oxide formation occurs therefore just at slightly higher oxygen chemical potentials than the first stabilization of surface configurations with adsorbed O atoms within the set of structures compared in our study (at $\Delta \mu_{\rm O} = -0.89$\,eV and $-$0.68\,eV respectively, cf. Figs. \ref{fig2} and \ref{fig3}). This is largely different compared to the situation at Pd(111), where a $p(2 \times 2)$ structure with 0.25\,ML on-surface O adatoms in fcc hollow sites gets stabilized at oxygen chemical potentials that are significantly lower than the onset of bulk oxide formation \cite{reuter04}. The reason for this difference is the high cost of segregating the scarce Pd in Ag$_3$Pd to the surface that is not necessary at Pd(111). The average O binding energy we compute at 0.25\,ML coverage at the Ag-rich surface termination (100\% Ag in the first layer, 100\% Pd in the second layer) that is most stable at the lowest $\Delta \mu_{\rm O}$ is only $-$0.13\,eV/O atom, which should be compared to our computed average binding energy of $-$1.29\,eV/O atom at the pure Pd(111) surface at this coverage. Only by segregating Pd atoms to the first layer (and thus involving them into the direct coordination to the O adsorbate), does the $E_{\rm b}$ of corresponding Ag$_3$Pd(111) configurations come into the same range as the Pd(111) binding energy, e.g. $-$1.29\,eV/O atom for the configuration with 100\% Pd in the first layer and 100\% Ag in the second layer, cf. Fig. \ref{fig5}. However, stabilizing such configurations then involves the cost of segregating Pd to the topmost layer and thus delays their stability range up to oxygen chemical potentials, where bulk oxide formation is already about to set in. 

Within the set of calculated $(2 \times 2)$ surface unit-cells, we can obviously not exclude that there are not more dilute Ag$_3$Pd(111) adsorption structures with $\theta < 0.25$\,ML that could already become most stable at a $\Delta \mu_{\rm O}$ that is noticeably lower than the limit of bulk oxidation. However, the essential physics should already be captured within our considered subset of configurations, and this is that Pd and Ag are chemically simply too similar, or more precisely that the O-Pd binding is not stronger by a sufficient amount compared to the O-Ag binding, to overcome the high Pd segregation cost in Ag$_3$Pd and induce a noticeable adsorbate-induced segregation at already lowest oxygen chemical potentials. Hence, segregation of Pd only sets in, when the driving force from the gas phase is already so high, that this directly initiates the formation of bulk-like oxide films at the surface. In this respect, it will be interesting to investigate how this compares to the situation in more Pd-rich Pd-Ag alloys, or to the case of binary alloys formed of more chemically different metals, like Pt-Ru alloys. There the strong O-Ru binding will exert a large driving force for Ru segregation to the topmost layer \cite{han05}, which on the other hand needs to overcome the much higher segregation energy to enrich the surface with the more reactive Ru metal \cite{ruban99a,ruban99b}. At which oxygen chemical potentials this then leads to the stabilization of Ru-rich O-containing surface structures, and how this $\Delta \mu_{\rm O}$ then compares to the onset of bulk oxidation, will depend sensitively on the availability of Ru atoms, and therewith on the detailed structure and composition of the bulk alloy reservoir.

\section{Summary}

In conclusion, we have presented a first-principles atomistic thermodynamics approach to describe the structure and composition of an alloy surface in contact with a (reactive) environment. Accounting for both the dependency on the bulk alloy and the gas phase reservoir, this approach provides the appropriate framework for a detailed discussion of the driving factors behind adsorbate-induced surface segregation. We illustrate this for the Ag$_3$Pd(111) surface exposed to an oxygen atmosphere, where we obtain an inversion of the segregation profile in increasingly oxygen-rich environments. Whereas a minimal segregation energy stabilizes Ag-terminated surface structures in ultra-high vacuum type gas phase conditions, the much stronger oxygen bonding favors increasingly Pd-rich terminations in atmospheres with higher oxygen content. Our analysis shows that the details of this transition are intricately coupled to the bulk reservoir structure, and already small non-stoichiometries in the nominally ordered Ag$_3$Pd bulk alloy can not only change the oxygen pressures required for the transition by several orders of magnitude, but also affect the explicit segregation profile itself. This highlights the restricted relevance of detailed segregation data from ultra-high vacuum experiments for applications in realistic environments, as well as the complexity that needs to be captured in the modeling to obtain quantitatively accurate results. In this respect our study shows that at least for the Ag$_3$Pd(111) surface simplified treatments based on impurity segregation and adsorption data of the parent metals are not able to correctly describe the adsorbate-induced segregation, nor that restricted information about the low-coverage adsorbate bonding at the alloy surface or the exclusive consideration of non-stoichiometries in the topmost substrate layer would be sufficient. This holds already for the case where the reactive environment leads only to the formation of on-surface adsorbate layers, and is even more pronounced when considering the possibility of a more complex restructuring like here the oxide formation in an O-containing environment. Also in the latter case the ruling factors are the availability of the more reactive metal species in the alloy bulk and the increased oxygen bond strength that this species can provide. In the studied case of the Ag$_3$Pd(111) surface the high cost of segregating the scarce Pd to the surface delays the stabilization of Pd-enriched surface structures in fact almost up to gas phase conditions where the formation of thicker palladium oxide films would already become thermodynamically favorable. We speculate that this will be different in more Pd-rich Ag-Pd alloys, or in alloys composed of species with a larger difference in their O binding capabilities.

\begin{acknowledgments}
JRK gratefully acknowledges the Alexander von Humboldt Foundation for financial support. JRK would like to thank Volker Blum for many discussions on alloys and thermodynamics, and Martin Fuchs for helpful discussions on pseudopotentials.
\end{acknowledgments}

\end{document}